\documentclass[12pt]{iopart}
\usepackage{graphicx}
\usepackage{caption}

\begin{document}

\title[ ]{FlexDTI: Flexible diffusion gradient encoding scheme-based highly efficient diffusion tensor imaging using deep learning }

\author{ $ Zejun Wu^{1}, Jiechao Wang^{1}, Zunquan Chen^{1}, Qinqin Yang^{1}, Zhen Xing^{2}, $}
\author{ $ Dairong Cao^{2}, Jianfeng Bao^{3}, Taishan Kang^{4}, Jianzhong Lin^{4}, $}
\author{ $ Shuhui Cai^{1}, Zhong Chen^{1}, Congbo Cai^{1*} $}
\author{ $  $}

\address{1. Department of Electronic Science, Fujian Provincial Key Laboratory of Plasma and Magnetic Resonance, Xiamen University, Xiamen 361005, China}
\address{2. Department of Radiology, The First Affiliated Hospital of Fujian Medical University, Taijiang District, Fuzhou 350005, China}
\address{3. Department of Magnetic Resonance Imaging, The First Affiliated Hospital of Zhengzhou University, Zhengzhou University, Zhengzhou 450052, China}
\address{4. Department of MRI, Zhongshan Hospital of Xiamen University, School of Medicine, Xiamen University, Xiamen 361004, China}
\address{* Corresponding author: Congbo Cai (e-mail: cbcai@xmu.edu.cn)}
\vspace{12pt}

\begin{abstract}
Objective: Most deep neural network-based diffusion tensor imaging methods require the diffusion gradients' number and directions in the data to be reconstructed to match those in the training data. This work aims to develop and evaluate a novel dynamic-convolution-based method called FlexDTI for highly efficient diffusion tensor reconstruction with flexible diffusion encoding gradient scheme.
Approach: FlexDTI was developed to achieve high-quality DTI parametric mapping with flexible number and directions of diffusion encoding gradients. The method used dynamic convolution kernels to embed diffusion gradient direction information into feature maps of the corresponding diffusion signal. Furthermore, it realized the generalization of a flexible number of diffusion gradient directions by setting the maximum number of input channels of the network. The network was trained and tested using datasets from the Human Connectome Project and local hospitals. Results from FlexDTI and other advanced tensor parameter estimation methods were compared. 
Main results: Compared to other methods, FlexDTI successfully achieves high-quality diffusion tensor-derived parameters even if the number and directions of diffusion encoding gradients change. It reduces normalized root mean squared error (NRMSE) by about 50$\%$ on fractional anisotropy (FA) and 15$\%$ on mean diffusivity (MD), compared with the state-of-the-art deep learning method with flexible diffusion encoding gradient scheme.
Significance: FlexDTI can well learn diffusion gradient direction information to achieve generalized DTI reconstruction with flexible diffusion gradient scheme. Both flexibility and reconstruction quality can be taken into account in this network.

\end{abstract}

%
\vspace{2pc}

\section{Introduction}

Diffusion weighted imaging (DWI) constitutes an essential part of magnetic resonance imaging modalities. It measures the extent of water molecule diffusion along the diffusion gradient direction (Tae \textit{et al} 2018). DWI can be used in the early detection of ischemic stroke (Edlow \textit{et al} 2017, Auriat \textit{et al} 2015, Moseley \textit{et al} 1990, Zhang \textit{et al} 2018) and brain tumors (Stadnik \textit{et al} 2001, Kono \textit{et al} 2001, Szczepankiewicz \textit{et al} 2015). Since diffusion can vary greatly along or perpendicular to the direction of the tissue microstructure, diffusion tensor is introduced to describe the anisotropy of tissue. Diffusion tensor imaging (DTI) has been shown to be useful for the detection of traumatic brain injury (Muller \textit{et al} 2016, Narayana \textit{et al} 2015, Pasternak \textit{et al} 2018), multiple sclerosis (Rovaris and Filippi 2007, Song \textit{et al} 2002, Van \textit{et al} 2013), and major depressive disorder (Elliott \textit{et al} 2011, Rive \textit{et al} 2013). In DTI, the diffusion properties in each voxel are described by a rank-2 symmetric diffusion tensor. Eigen decomposition is applied to the diffusion tensor to get DTI parametric maps such as fractional anisotropy (FA) and mean diffusivity (MD), which can visualize the tissue microstructures. There exist more sophisticated diffusion techniques, such as diffusion kurtosis imaging (Li \textit{et al} 2019) and high angular resolution diffusion imaging (Zhang \textit{et al} 2012), but DTI remains a widely used tool for neuroscience research (Tetreault \textit{et al} 2020).\\ 
The reconstruction of DTI theoretically requires a minimum of six DW images acquired along non-collinear diffusion gradient directions, as well as one non-DW image. These images are processed using the linear least square (LLS) fitting algorithm. However, high-quality DTI usually requires more DW images due to the low signal-to-noise ratio (SNR) of DWI, which causes a long scan time. For instance, it is typically necessary to acquire 15-20 DW images to generate DTI parametric maps with sufficient diagnostic quality (Jones \textit{et al} 2004, Skare \textit{et al} 2000), leading to a scan time of 4-5 minutes. Such a long scan time would make the patient uncomfortable, and increases the chance of motion artifacts corruption.\\ 
Recently, deep neural networks have been widely applied in Magnetic Resonance Imaging (MRI) reconstruction, such as in compressed sensing (Liu \textit{et al} 2022, Wang \textit{et al} 2023, Zhang \textit{et al} 2020), $T_2$ mapping (Zhang \textit{et al} 2019, Cai \textit{et al} 2018, Yang \textit{et al} 2022), and DTI (Golkov \textit{et al} 2016, Tian \textit{et al} 2020, Li \textit{et al} 2021). The q-space deep learning (q-DL) (Golkov \textit{et al} 2016) is one of the earliest deep learning methods for diffusion parameter estimation, including diffusion kurtosis as well as neurite orientation dispersion and density measures. The results have shown that a simple three-layer neural network can accurately estimate diffusion parameters and shorten scan time by twelve-fold due to the reduction of required diffusion gradient directions. Thereafter, several deep learning-based methods (e.g., DeepDTI (Tian \textit{et al} 2020), and SuperDTI (Li \textit{et al} 2021)), which use only six-direction DW images, were proposed to further shorten scan time.\\ 
Even though the present deep learning methods can reconstruct high-quality diffusion tensor with a small number of diffusion gradient directions, they require that the number and direction of the diffusion gradients of the data to be reconstructed must be the same as the training data. For different diffusion gradient schemes, the networks need to be retrained, and a large amount of training data is additionally required. This issue of generalization hinders the practical applications of deep learning methods since different diffusion encoding schemes are often used in MRI scanners from different manufacturers such as Siemens and Philips. Although the traditional LLS fitting method has good generalization performance, the accuracy of fitting results is not high when the number of diffusion gradient directions is few. To solve this issue, a new method called DIFFnet (Park \textit{et al} 2022) was proposed. In this method, diffusion signals were normalized in the q-space and then projected and quantized in an orthogonal plane, producing a matrix as an input for the network to achieve generalization. Nevertheless, when the number of diffusion gradient directions is reduced, the projection will cause the signal to appear sparse and scattered on the orthogonal plane, reducing the quality of DTI.\\ 
To build a flexible DTI reconstruction network, diffusion gradient direction information needs to be efficiently input into the network, because there are significant differences in the DW images with different diffusion gradient directions. Chen \textit{et al}. (2020) designed a new convolutional kernel that dynamically aggregates multiple parallel convolution kernels based on input dependence. Zhang \textit{et al}. (2021) utilized dynamic convolutional kernels to achieve the fusion of classification tasks and input information. Huang \textit{et al}. (2023) used dynamic convolution to fuse spatial and physical information with different dimensions and overcome the receptive field limitation of the convolutional network. In this work, we implement a dynamic-convolution-based network to get the dynamical kernel parameters conditioned by the DW images and diffusion gradient direction information. Furthermore, our method realizes the generalization of flexible number of diffusion gradient directions by setting the maximum number of input channels of the network. Compared to the conventional fitting method and DIFFnet, the results of the proposed method have a lower normalized root mean squared error (NRMSE) and a higher peak signal-to-noise ratio (PSNR) and structure similarity index measure (SSIM) in the DTI parametric maps. 

\section{Methods}

\subsection{DTI model}
The voxel-wise magnitude signal in DTI is described by the Stejskal-Tanner equation (Stejskal \textit{et al} 1965): 
\begin{eqnarray}
S_{i} = S_{0}exp(-b\mathit{\mathbf{g}}_{i}^{T}\mathit{\mathbf{D}}\mathit{\mathbf{g}}_{i}) \label{eq1}
\end{eqnarray}
in which \textit{$S_i$} is the signal intensity for a special \textit{b} value of diffusion (\textit{$S_0$} corresponding to \textit{b} = 0). \textit{$\textbf{g}_i=(g_{ix}, g_{iy}, g_{iz})^T$} is the unit direction vector of diffusion, and diffusion tensor (\textbf{D}) is:
\begin{eqnarray}
\mathit{\mathbf{D}} = 
\left(\begin{array}{ccc}
  D_{xx} & D_{xy} & D_{xz} \\
  D_{xy} & D_{yy} & D_{yz} \\
  D_{xz} & D_{yz} & D_{zz}
\end{array}
\right)
\end{eqnarray}

\noindent
By matrix linear transformation 
\textit{${\alpha}_i = (g_{ix}^2, g_{iy}^2, g_{iz}^2, 2g_{ix}g_{iy}, 2g_{ix}g_{iz}, 2g_{iy}g_{iz})^T$}, the diffusion tensor element {${D}_{vec} = (D_{xx}, D_{yy}, D_{zz}, D_{xy}, D_{xz},D_{yz})^T$} and the apparent diffusion coefficients  \textit{${\beta}_i = ln(S_0/S_i)/b$} can be expressed as:
\begin{eqnarray}
\mathbf{{\alpha}}_{i}^{T} {D}_{vec} = {\beta} _{i} 
\end{eqnarray}
The diffusion tensor elements \textbf{\textit{${D}_{vec}$}} consist of six variables, which mathematically requires at least six \textit{$S_i$} of non-collinear directions for the LLS method.

\noindent
Then, the eigenvalues ($ {\lambda}_{1}, {\lambda}_{2}, {\lambda}_{3} $) and corresponding eigenvectors ($ {V}_{1}, {V}_{2}, {V}_{3} $) can be acquired through eigen decomposition of the diffusion tensor $ {D}_{vec} $. Based on the obtained eigenvalues, DTI parameters, including FA, MD, Axial Diffusivity (AD), and Radial Diffusivity (RD), can be calculated using the formulas (4-7).

\begin{eqnarray}
FA =  \frac{\sqrt{3}}{\sqrt{2}}  \frac{\sqrt{({\lambda}_{1}-\lambda)^2+({\lambda}_{2}-\lambda)^2+({\lambda}_{3}-\lambda)^2}}{\sqrt{\lambda_{1}^2+\lambda_{2}^2+\lambda_{3}^2}}
\end{eqnarray}

\begin{eqnarray}
MD =  \frac{1}{3}(\lambda_{1}+\lambda_{2}+\lambda_{3})
\end{eqnarray}

\begin{eqnarray}
AD =  \lambda_{1}
\end{eqnarray}

\begin{eqnarray}
RD =  \frac{1}{2}(\lambda_{2}+\lambda_{3})
\end{eqnarray}

\noindent
The FA values serve as criteria for tissue anisotropy and integrity, while MD values quantify the diffusion degree of water molecules per voxel. Additionally, the AD and RD values represent the eigenvalues corresponding to water diffusion along and perpendicular to the long axis of the fiber bundle, respectively.

\subsection{Data acquisition}
The data were obtained from the Human Connectome Project (HCP) WU-Minn-Ox Consortium public database (https://www.humanconnectome.org), which was approved by the institutional research ethics committees (Van Essen \textit{et al} 2012). There were 203 healthy subjects in total, among which 136 subjects were used for training, 46 subjects for validation, and 21 subjects for testing. DW images with two \textit{b} values (\textit{b} = 0, 1000 s/$mm^2$) and 90 diffusion gradient directions were used. 90 diffusion gradient directions were acquired in a way that maintains a global uniform angular coverage always (Caruyer \textit{et al} 2013). The MRI data were acquired on a 3T MRI scanner (CONNECTOME Skyra, Siemens, Germany). The parameters of the diffusion-weighted spin echo Echo-Planar Imaging (EPI) sequence were as follows: TR = 5,520 ms, TE = 89.4 ms, slice thickness = 1.25 mm, FOV = 210×180$mm^2 $, matrix size = 168×144, slice number = 111, pulse flip angle = 78°, multiband factor = 3, echo spacing = 0.78 ms and phase partial Fourier = 6/8. The $T_1$-weighted images were acquired using a 3D MPRAGE sequence with the following parameters: TR = 2,400 ms, TE = 1.87 ms, slice thickness = 0.7 mm, FOV = 224×224 $ mm^2 $, matrix size = 168×144, slice number = 256, and pulse flip angle = 8°.\\
Six healthy subjects were also included for testing and analysis. Three of the healthy subjects were recruited for data collection at Zhejiang University, where MRI data, including DW images, $T_1$-weighted and $T_2$-weighted images, were acquired using a 3T MRI scanner (MAGNETOM Prisma, Siemens, Germany). The sequence parameters were the same as those used for HCP data except for the diffusion gradient direction scheme, which underwent certain changes compared to the HCP’s gradient direction scheme. DW images with 90 diffusion gradient directions (\textit{b} = 1000 s/$mm^2$) and two \textit{b} = 0 images with reversed phase encoding directions were acquired. The remaining three healthy subjects were recruited for data collection at Zhongshan Hospital, Xiamen, where MRI data was acquired on a 3T MRI scanner (Ingenia CX, Philips Healthcare, Best, The Netherlands) with a typical 2×2×2 $mm^3$ acquisition. The study protocol was approved by the local institutional research ethics committee, and written informed consents were obtained from the volunteers before the experiments.\\
We applied the diffusion preprocessing pipeline from the minimal preprocessing pipelines (Glasser \textit{et al} 2013) for MRI data, which started by normalizing the intensities of the mean \textit{b} = 0 images. These pairs of \textit{b} = 0 images with reversed phase encoding directions were utilized to estimate and correct the EPI distortion (Jovicich \textit{et al} 2006). Subsequently, the estimated distortion field was input into a Gaussian process predictor, which collectively utilized all the available data to estimate the eddy-current induced field inhomogeneity and head motion for each volumetric image (Andersson \textit{et al} 2003). The diffusion data underwent the calculation of the warped field for gradient nonlinearity correction, reducing spatial distortions. The gradient nonlinearity corrected image was registered to the $T_1$-weighted structural image using the FLIRT BBR cost function (FSL5) and the FreeSurfer's boundary-based registration for fine-tuning (Sotiropoulos \textit{et al} 2013). 

\subsection{Dynamic convolutional network}
The input of the network shown in Figure 1 were: a \textit{b} = 0 image and different numbers of DW images of \textit{b} = 1000 s/$mm^2$ along different diffusion gradient directions were randomly selected from the first 50 directions in the 90 directions of HCP for training and from the last 40 directions for testing. \\
We implemented a dynamic-convolution-based model to achieve generalized reconstruction for various diffusion gradient directions. Firstly, each DW image with \textit{b} = 1000 s/$mm^2$ was separately fused with diffusion gradient features through a dynamic convolution module, obtaining a DW feature map with diffusion gradient encoding information. Here, U-Net (Falk \textit{et al} 2019) was used to achieve direct mapping from DW feature maps to diffusion tensors. In previous DTI networks, the diffusion gradient encoding scheme was generally fixed, which limited their flexibility. In this work, FlexDTI achieved the generalization with the flexible number of DW images by setting the maximum number of input channels. After the input DW image passed through the dynamic convolution module, the obtained feature map and \textit{b} = 0 image were overlaid as the input to U-Net. Assuming that the maximum number of input channels for U-Net is N, when the number of channels was less than N, the remaining channels would be filled with duplicate feature maps. This is a method to handle situations where the number of \textit{b}-values is variable, ensuring the proper functioning of the network. The output of the network was six different elements ($ D_{xx}, D_{yy}, D_{zz}, D_{xy}, D_{xz},D_{yz} $) of the diffusion tensor. The reconstructed diffusion tensor was then through Eigen decomposition to obtain the DTI quantitative maps, including FA, MD, AD and RD.\\
The implementation of the dynamic convolutional module is demonstrated in Figure 1(b). The DW images ($\textbf{W}$) were aggregated into a one-dimensional vector and were connected with a one-dimensional diffusion gradient direction vector ($\textbf{V}$) via global average pooling (GAP), which had a global receptive field and realized information fusion between diffusion encoding direction and corresponding DW images. By specifying the parameters of dynamic convolution layers ($\theta$), the concatenated vector through the attention mechanism module ($\psi$) was utilized to get the kernel parameter. This process was expressed as follows:
\begin{eqnarray}
\omega = {\psi} (GAP(\mathbf{W})||\mathbf{V};{\theta}_{\psi})
\end{eqnarray}
where ${\theta}_{\psi}$ represented the weights and the biases of the dynamic convolution layers. The kernel parameter $\omega$ was allocated to three dynamic convolution layers. Within the dynamic convolutional module, the length of the one-dimensional DWI vector was ten through global average pooling, and the length of the diffusion gradient direction vector is three. Under the condition of 3×3 convolution kernel, the weights and biases for the three layers of dynamic convolution were set as follows: the first layer had 27 weights and 3 biases, the second layer had 81 weights and 3 biases, and the third layer had 81 weights and 3 biases, and a total of 198 parameters were generated.\\
The U-Net network structure was composed of 3×3 convolutional blocks repeatedly. Each block included a ReLU (Rectified Linear Unit) activation and a 2×2 maximum pooling operation. Given feature maps ${W}_{fm}$ formed by passing DW images through the dynamic convolution module, the output images were:
\begin{eqnarray}
D = f({W}_{fm};{\theta}_{f})
\end{eqnarray}

\noindent
where {${\theta}_{f}$} represented all network parameters to be learned during training.

\noindent
The sum of the average mean squared error ({$L_2$}) between final output image {$\overline{D_{j}}$} and the corresponding ground-truth diffusion tensor image {$D_{j}$} is defined as the loss function:
\begin{eqnarray}
Loss = ||\overline{D_{j}} -  D_{j}||_{2}^{2} 
\end{eqnarray}

\noindent
The training of DIFFnet was performed on NVIDIA GTX 1080Ti using TensorFlow. The training process took about 10 hours and was stopped after 50 epochs. The initial learning rate was 0.001 with a decaying factor of 0.87 for each epoch and the batch size was 100. We followed some of the hyperparameters used in DIFFnet and made adjustments based on our specific circumstances. FlexDTI took about 10 hours for training and about 147.3 ms per slice for testing. The initial learning rate was 0.001, with a decay applied every 80 epochs using a multiplicative factor of 0.1. The training process was stopped after 100 epochs and the batch size was 16.

\begin{figure*}[ht]
\includegraphics[width=35pc,height=30pc]{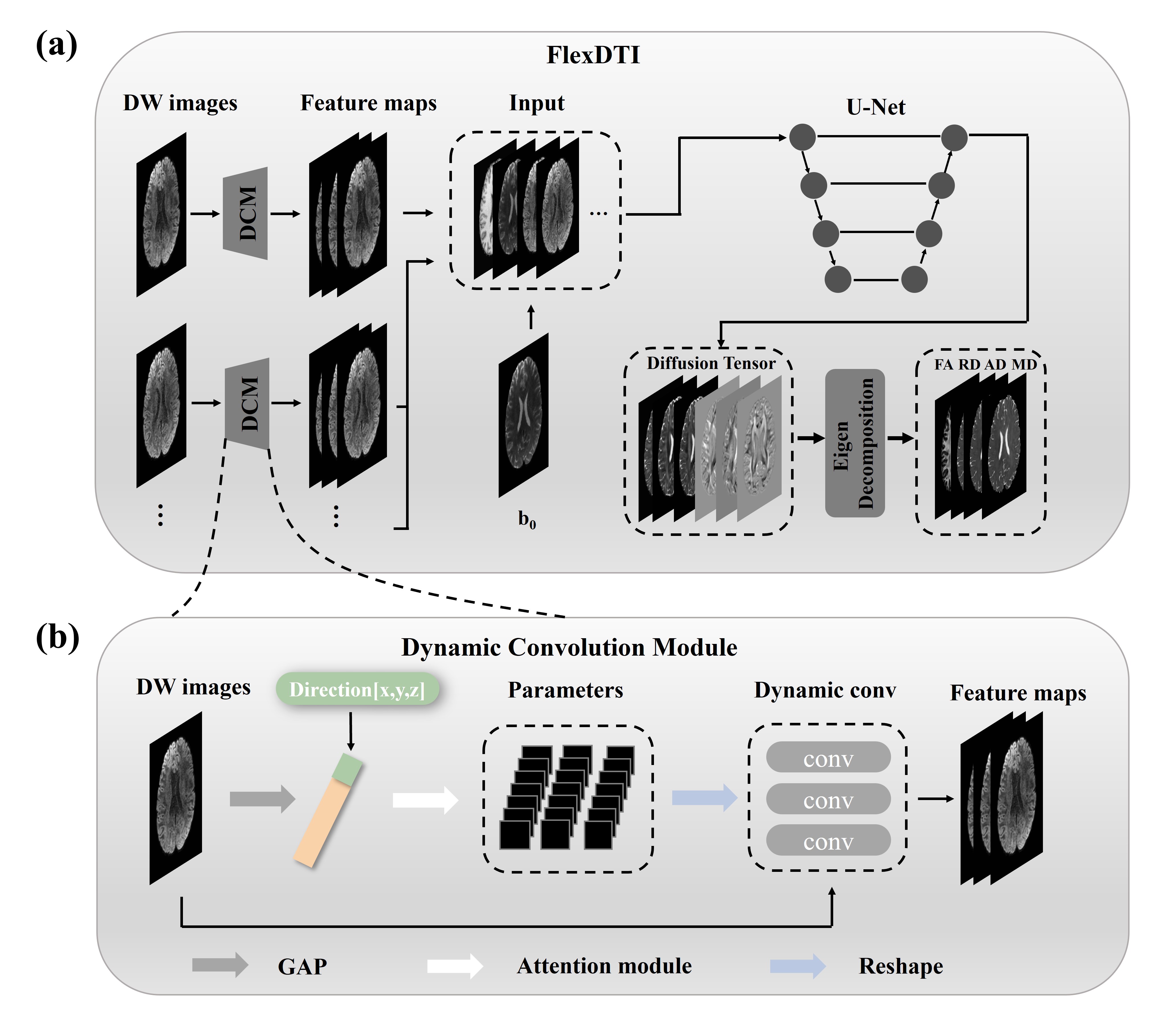}
\centering
\caption{(a) Overall architecture of the proposed method for generalized DTI reconstruction. The input of the network has one \textit{b} = 0 image and six DW images from six diffusion directions. The output of the network is diffusion tensor, and Eigen decomposition is used to fit the DTI parametric maps. (b) The dynamic convolution module to achieve generalized reconstruction of DTI. The DWIs are aggregated into a one-dimensional vector via GAP concatenated with the diffusion direction vector. The concatenated vector is convolved and fully connected to get the kernel parameters, which are allocated to three dynamic convolution layers.}\label{Figure 1}
\end{figure*}

\subsection{Evaluation}
The one \textit{b} = 0 image and 90 DW images of \textit{b} = 1000 s/$mm^2$ were used to obtain the ground truth based on the Stejskal-Tanner equation. \\
To demonstrate the effectiveness of the proposed method, a series of comparative experiments were carried out, which all supported flexible diffusion gradient directions, including conventional LLS, LLS after block-matching and 4D filtering (LLS-BM4D) (Dabov \textit{et al} 2007) and DIFFnet methods. SuperDTI (fixed diffusion gradient scheme) was also performed for comparison. There were two sets of validation experiments, one was under fixed diffusion gradient number and flexible diffusion gradient directions, and the other was under flexible number and directions of diffusion gradients. For the first set of experiments, the diffusion gradient directions number was fixed at 6 and 12 respectively. For the second set of experiments, a maximum of 20 diffusion gradient directions was set. \\
The PSNR, SSIM and NRMSE were calculated to quantify the reconstruction quality. The PSNR is defined as follows: 
\begin{eqnarray}
PSNR = 10\log_{10}{\frac{Peak^{2}}{\sum_{i=1}^{V}[\overline{v(i)} - v_{ref}(i)]^{2}/ V}} 
\end{eqnarray}
where \textit{Peak} is the maximum value of all pixel values in the reference image, $\overline{v}$ is the reconstructed image and $v_{ref}$ is the reference image. while the peak value is typically associated with specific regions or features, such as the pixel with the highest FA in white matter or the highest MD in CSF, the purpose of PSNR is to comprehensively evaluate the overall distortion of the entire image.\\
The SSIM is defined as follows: 
\begin{eqnarray}
SSIM(v_{ref}, \overline{v}) = \frac{(2{\mu}_{\overline{v}}{\mu}_{v_{ref}}+c_1)(2{\sigma}_{\overline{v},v_{ref}}+c_2)}{({{\mu}_{\overline{v}}}^2+{{\mu}_{v_{ref}}^2+c_1)({{\sigma}_{\overline{v}}}^2+{{\sigma}_{v_{ref}}^2}+c_2)}}
\end{eqnarray}

\noindent
where ${\mu}_{\overline{v}}$ and ${\mu}_{v_{ref}}$ are averaged signal intensities, ${{\sigma}_{\overline{v}}}^2$ and ${\sigma}_{v_{ref}}^2$ are the standard deviations of $\overline{v}$ and $v_{ref}$, ${\sigma}_{\overline{v},v_{ref}}$ is the covariance of $\overline{v}$ and $v_{ref}$, and $c_1$ and $c_2$ are two offset constants to stabilize the division.
NRMSE is defined as follows:
\begin{eqnarray}
NRMSE = \sqrt{\frac{\sum_{i=1}^{V}[\overline{v(i)} - v_{ref}(i)]^{2}}{\sum_{i=1}^{V}(v_{ref}(i))^{2}}}
\end{eqnarray}
NRMSE normalizes the root mean square error (RMSE) by considering the maximum pixel value in the reference image, thereby accounting for the scale of the reference image.

\section{Results}

\subsection{DTI reconstruction with flexible diffusion gradient directions}
The results of four estimated tensor-derived parameters, i.e., FA, MD, AD and RD through different reconstruction methods were provided in Figure 2, where 6 DW images with flexible diffusion gradient directions were involved in the reconstruction. In terms of the reconstruction quality of the four quantitative parameters, FlexDTI has significant performance improvement compared with the other methods. The images from FlexDTI appear less noisy than others, with details more similar to the reference images.  It is worth pointing out that 6 DW images are not sufficient to reconstruct high-quality DTI for DIFFnet method, while the proposed method can achieve high-quality DTI quantitative maps using only 6 DW images with flexible diffusion gradient directions.

\begin{figure*}[ht]
\includegraphics[width=30pc,height=48pc]{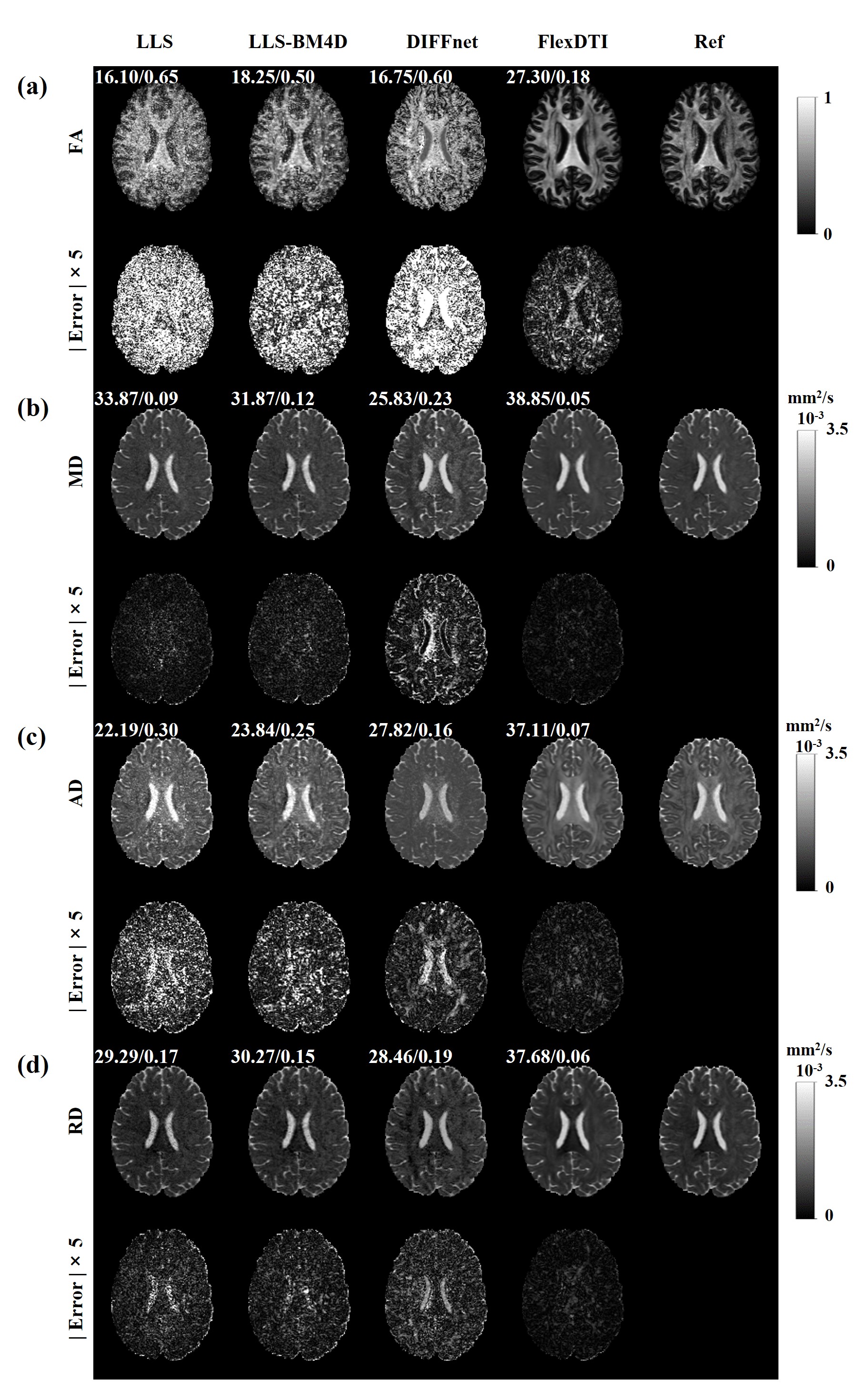}
\centering
\caption{Four parametric maps (FA, MD, AD and RD) reconstructed by LLS, LLS-BM4D, DIFFnet and FlexDTI using 6 DW images with flexible diffusion gradient directions. The references were reconstructed by LLS using 90 DW images. The PSNRs and NRMSEs are given at the upper left corner of each reconstruction image.}\label{Figure 2}
\end{figure*}

\noindent
Table 1 presents the mean ± standard deviation of PSNR, SSIM and NRMSE for the data collected from 21 volunteers. These metrics correspond to the four estimated tensor-derived parameters obtained through four different methods. Compared to DIFFnet, FlexDTI reduces NRMSE by 0.54, increases PSNR by 8.15 dB and SSIM by 0.17 on FA, and reduces NRMSE by 0.17, increases PSNR by 10.36 dB and SSIM by 0.05 on MD.

\begin{table*}[ht]
\caption{Quantitative assessment of FA, MD, AD and RD obtained with different methods using 6 DW images from HCP dataset.}
\begin{tabular}{cccccc}
\hline
\textbf{DTI parametric map} & \textbf{Metrics} & \textbf{LLS} & \textbf{LLS-BM4D} & \textbf{DIFFnet} & \textbf{FlexDTI} \\ \hline
\textbf{{FA}}              & PSNR (dB)        & 17.90 ± 2.71        & 19.98 ± 2.79             & 20.35 ± 2.27            & \textbf{28.50 ± 2.32}   \\
                           & SSIM             & 0.79 ± 0.11        & 0.82 ± 0.09             & 0.77 ± 0.12            & \textbf{0.94 ± 0.03}   \\
                           & NRMSE            & 1.01 ± 0.51         & 0.79 ± 0.35              & 0.80 ± 0.24             & \textbf{0.26 ± 0.10}    \\ \hline
\textbf{{MD}}              & PSNR (dB)        & 35.89 ± 2.86        & 33.32 ± 2.22             & 28.59 ± 2.25            & \textbf{38.95 ± 1.43}   \\
                           & SSIM             & 0.94 ± 0.02        & 0.94 ± 0.03             & 0.93 ± 0.03            & \textbf{0.98 ± 0.01}   \\
                           & NRMSE            & 0.11 ± 0.04         & 0.16 ± 0.22              & 0.25 ± 0.03             & \textbf{0.08 ± 0.04}    \\ \hline
\textbf{{AD}}              & PSNR (dB)        & 24.91 ± 2.93        & 26.24 ± 2.49             & 30.80 ± 2.33            & \textbf{36.61 ± 1.85}   \\
                           & SSIM             & 0.88 ± 0.07        & 0.90 ± 0.05             & 0.92 ± 0.04            & \textbf{0.97 ± 0.01}   \\
                           & NRMSE            & 0.36 ± 0.11         & 0.32 ± 0.23              & 0.18 ± 0.04             & \textbf{0.09 ± 0.04}    \\ \hline
\textbf{{RD}}              & PSNR (dB)        & 31.64 ± 2.61        & 32.07 ± 2.36             & 26.02 ± 2.23            & \textbf{38.11 ± 1.52}   \\
                           & SSIM             & 0.94 ± 0.03        & 0.94 ± 0.03             & 0.90 ± 0.05            & \textbf{0.98 ± 0.01}   \\
                           & NRMSE            & 0.19 ± 0.05         & 0.19 ± 0.26              & 0.35 ± 0.05             & \textbf{0.09 ± 0.06}    \\ \hline
                            
\end{tabular}
\end{table*}

\noindent
In Figure 3, 12 DW images with flexible diffusion gradient directions were used. Similar to the results with 6 DW images, FlexDTI has significant performance improvement compared to the other methods. Table 2 presents the mean ± standard deviation of PSNR, SSIM and NRMSE. Compared to DIFFnet, FlexDTI reduces NRMSE by 0.15, increases PSNR by 4.12 dB and SSIM by 0.05 on FA, and reduces NRMSE by 0.13, increases PSNR by 8.85 dB and SSIM by 0.02 on MD.  

\begin{figure*}[ht]
\includegraphics[width=30pc,height=48pc]{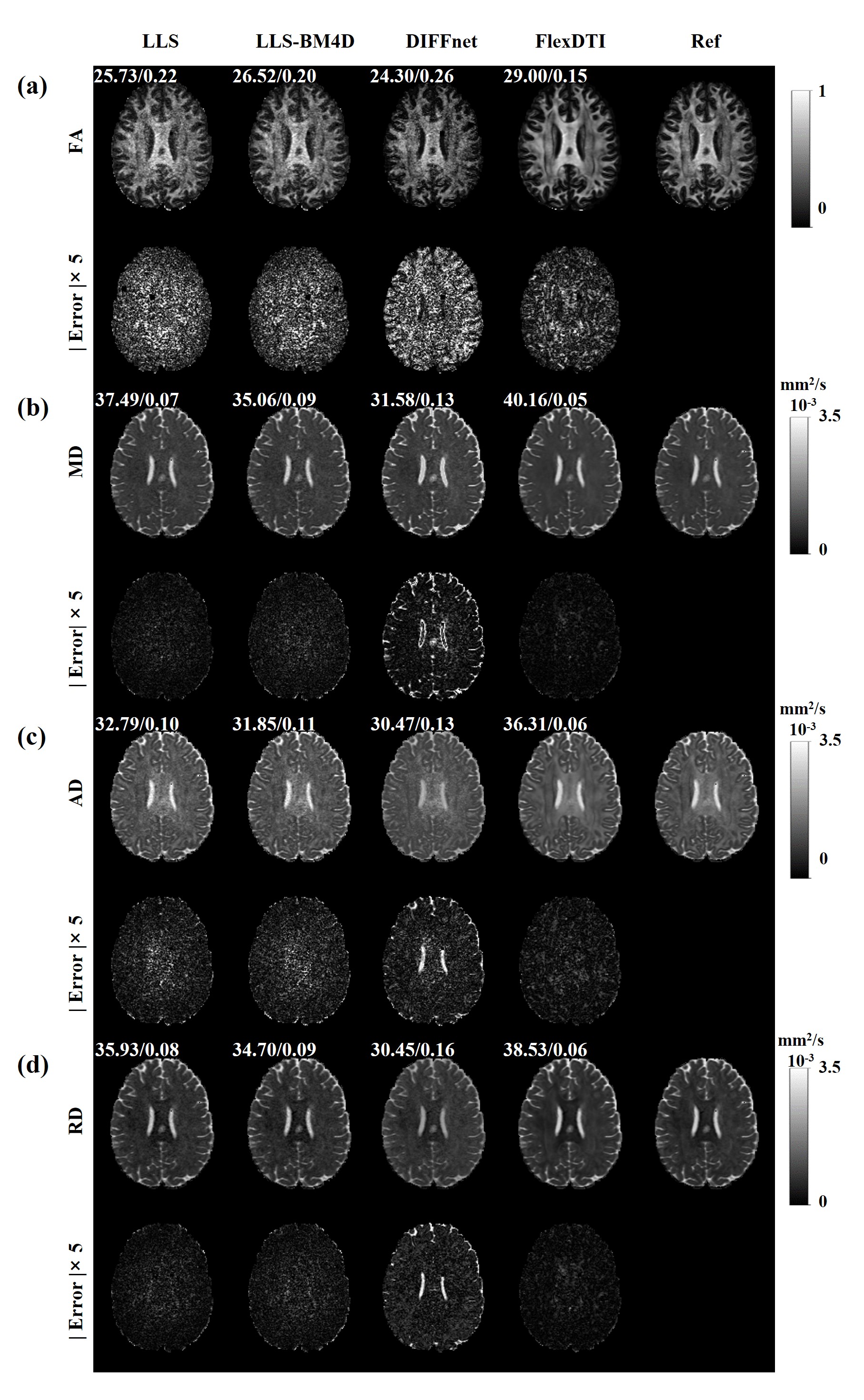}
\centering
\caption{Four parametric maps (FA, MD, AD and RD) reconstructed by LLS, LLS-BM4D, DIFFnet and FlexDTI using 12 DW images with flexible diffusion gradient directions. The references were reconstructed by LLS using 90 DW images. The PSNRs and NRMSEs are given at the upper left corner of each reconstruction image.}\label{Figure 3}
\end{figure*}

\begin{table*}[ht]
\caption{Quantitative assessment of FA, MD, AD and RD obtained with different methods using 12 DW images from HCP dataset.}
\begin{tabular}{cccccc}
\hline
\textbf{DTI parametric map} & \textbf{Metrics} & \textbf{LLS} & \textbf{LLS-BM4D} & \textbf{DIFFnet} & \textbf{FlexDTI} \\ \hline
\textbf{{FA}}              & PSNR (dB)        & 25.40 ± 2.85        & 26.28 ± 2.88             & 25.35 ± 2.03            & \textbf{29.47 ± 1.94}   \\
                           & SSIM             & 0.91 ± 0.05        & 0.91 ± 0.05             & 0.89 ± 0.05            & \textbf{0.94 ± 0.03}   \\
                           & NRMSE            & 0.43 ± 0.23         & 0.39 ± 0.22              & 0.41 ± 0.16             & \textbf{0.26 ± 0.13}    \\ \hline
\textbf{{MD}}              & PSNR (dB)        & 37.76 ± 2.89        & 35.62 ± 2.48             & 30.62 ± 2.34            & \textbf{39.47 ± 1.34}   \\
                           & SSIM             & 0.97 ± 0.02        & 0.96 ± 0.02             & 0.96 ± 0.02            & \textbf{0.98 ± 0.01}   \\
                           & NRMSE            & 0.09 ± 0.04         & 0.13 ± 0.21              & 0.20 ± 0.03             & \textbf{0.07 ± 0.04}    \\ \hline
\textbf{{AD}}              & PSNR (dB)        & 32.88 ± 2.70        & 32.18 ± 2.52             & 29.58 ± 2.02            & \textbf{36.88 ± 1.81}   \\
                           & SSIM             & 0.95 ± 0.02        & 0.95 ± 0.03             & 0.94 ± 0.03            & \textbf{0.97 ± 0.01}   \\
                           & NRMSE            & 0.14 ± 0.05         & 0.16 ± 0.09              & 0.20 ± 0.03             & \textbf{0.09 ± 0.03}    \\ \hline
\textbf{{RD}}              & PSNR (dB)        & 36.47 ± 3.00        & 35.35 ± 2.58             & 29.97 ± 2.18            & \textbf{38.53 ± 1.25}   \\
                           & SSIM             & 0.96 ± 0.02        & 0.96 ± 0.02             & 0.95 ± 0.02            & \textbf{0.98 ± 0.01}   \\
                           & NRMSE            & 0.11 ± 0.05         & 0.14 ± 0.02              & 0.22 ± 0.03             & \textbf{0.09 ± 0.07}    \\ \hline
                            
\end{tabular}
\end{table*}

\subsection{DTI reconstruction with flexible number and directions of diffusion gradients}

To validate the performance of reconstruction using flexible diffusion gradient direction number (6 or more), 6 to 20 DW images with flexible diffusion gradient directions were randomly selected from the first 50 diffusion gradient directions of HCP for training, and 6, 8, 12 and 20 diffusion gradient directions were selected from the last 40 diffusion gradient directions of HCP for testing.\\
Figure 4 shows that as the number of diffusion gradient directions increases, both FlexDTI and DIFFnet exhibit an improvement in the reconstruction quality of FA and MD, while FlexDTI consistently demonstrates significantly higher reconstruction quality than DIFFnet method. 

\begin{figure*}[ht]
\includegraphics[width=38pc,height=22pc]{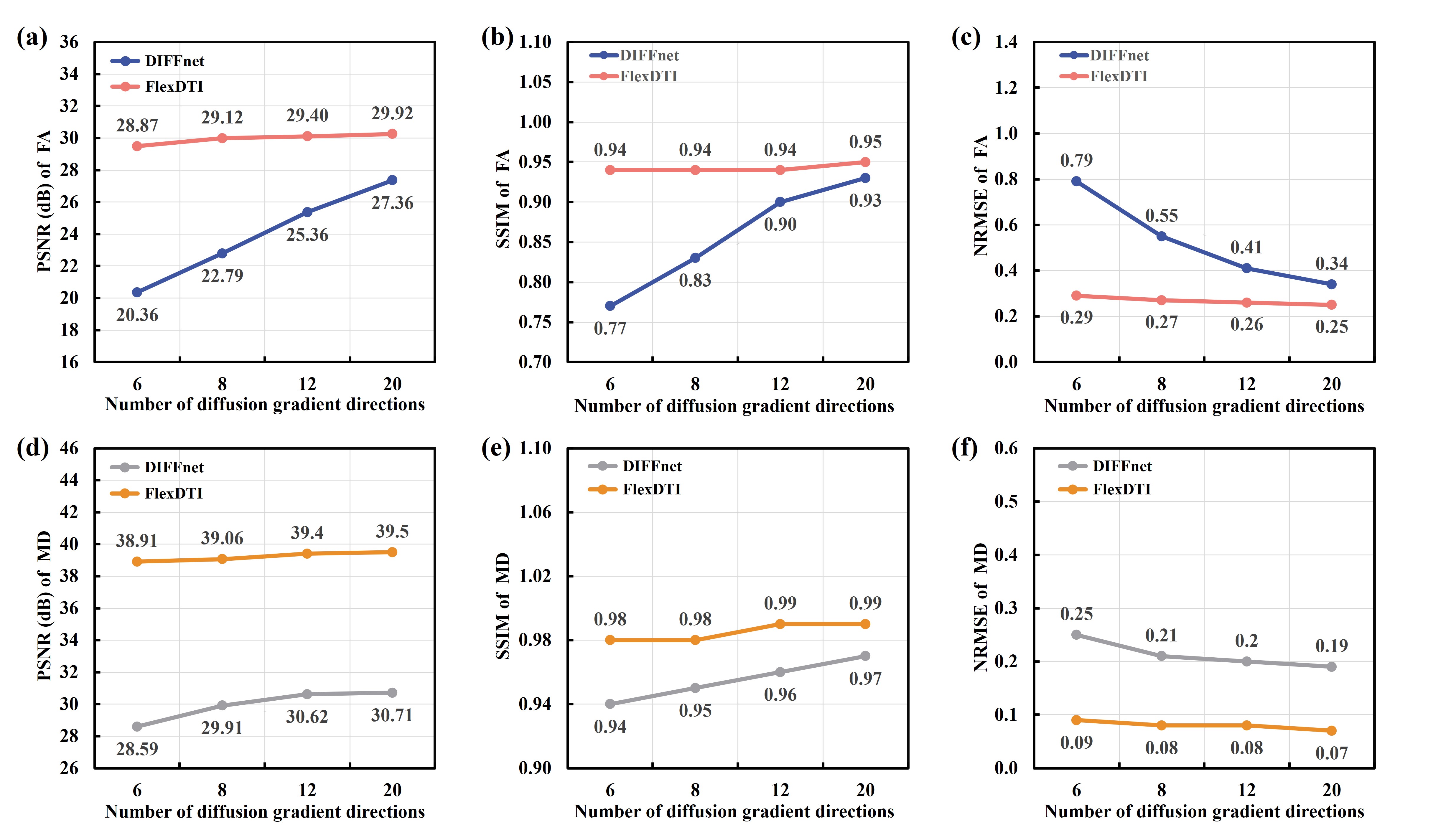}
\centering
\caption{The reconstruction quality assessment of FA and MD under different diffusion gradient direction numbers.}\label{Figure 4}
\end{figure*}

\subsection{DTI reconstruction of local clinical data}

Figure 5 and Table 3 show the results of further evaluations of our method on 3 test subjects from the local hospital volunteers. We used 6 DW images with flexible diffusion gradient directions to test the reconstruction effects of tensor-derived variables from the four methods. It can be observed that our method is still able to reconstruct better results than the other three methods, as demonstrated in Figure 5 and Table 3.

\begin{figure*}[ht]
\includegraphics[width=30pc,height=48pc]{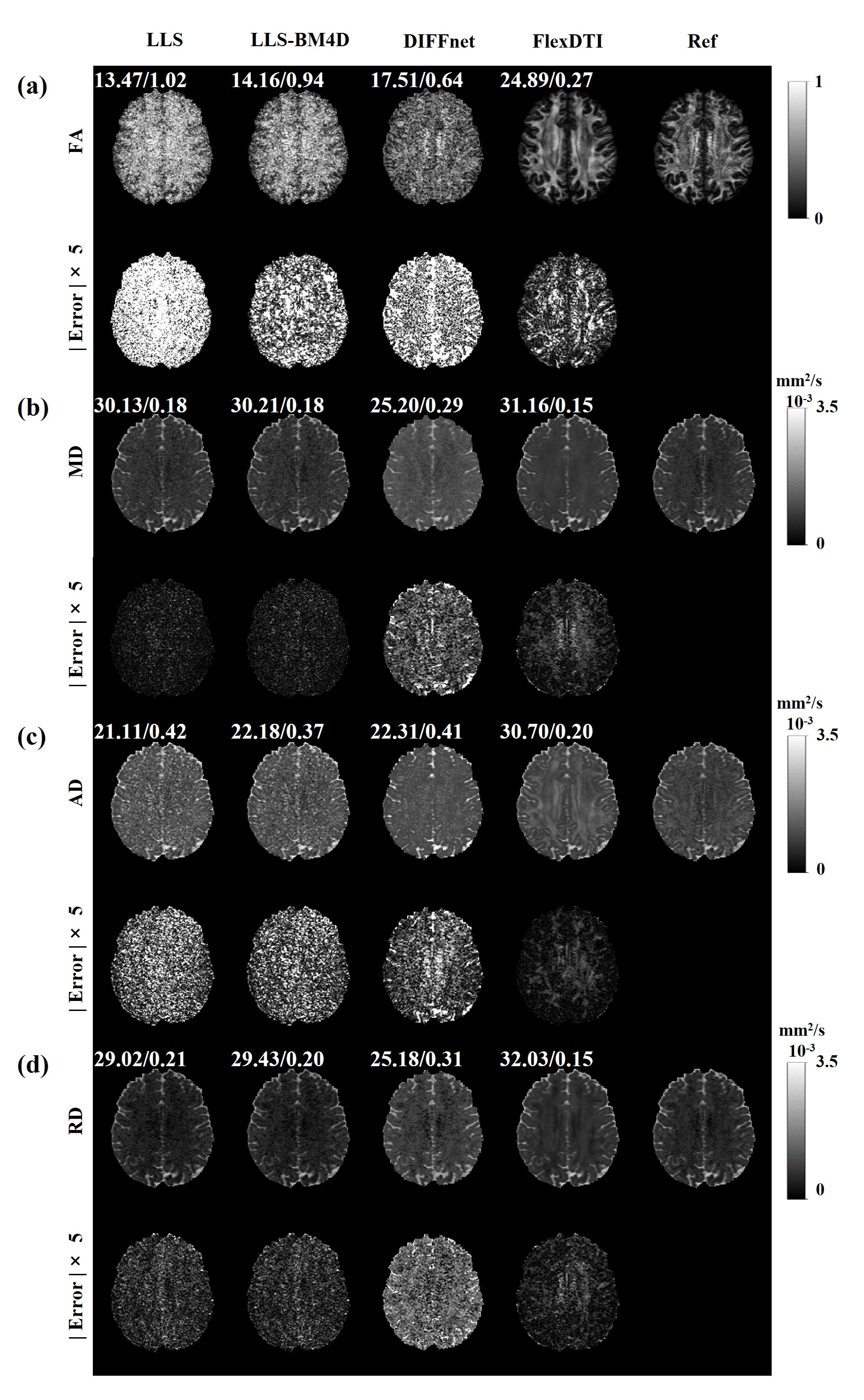}
\centering
\caption{Four parametric maps (FA, MD, AD and RD) reconstructed by LLS, LLS-BM4D, DIFFnet and FlexDTI using 6 DW images with flexible diffusion gradient directions for the local dataset. The references were reconstructed by LLS using 90 DW images. The PSNRs and NRMSEs are given at the upper left corner of each reconstruction image. }\label{Figure 5}
\end{figure*}

\begin{table*}[ht]
\caption{Quantitative assessment of FA, MD, AD and RD obtained with different methods using 6 DW images from local dataset.}
\begin{tabular}{cccccc}
\hline
\textbf{DTI parametric map} & \textbf{Metrics} & \textbf{LLS} & \textbf{LLS-BM4D} & \textbf{DIFFnet} & \textbf{FlexDTI} \\ \hline
\textbf{{FA}}              & PSNR (dB)        & 14.98 ± 2.25        & 15.68 ± 2.32             & 17.75 ± 2.41            & \textbf{26.03 ± 2.17}   \\
                           & SSIM             & 0.74 ± 0.11        & 0.75 ± 0.11             & 0.76 ± 0.10            & \textbf{0.90 ± 0.04}   \\
                           & NRMSE            & 1.35 ± 0.59         & 1.24 ± 0.50              & 0.76 ± 0.16             & \textbf{0.36 ± 0.06}    \\ \hline
\textbf{{MD}}              & PSNR (dB)        & 30.08 ± 2.15        & 29.28 ± 2.66             & 25.79 ± 2.99            & \textbf{30.17 ± 1.89}   \\
                           & SSIM             & 0.94 ± 0.02        & 0.94 ± 0.02             & 0.93 ± 0.04            & \textbf{0.94 ± 0.02}   \\
                           & NRMSE            & 0.20 ± 0.02         & 0.10 ± 0.02              & 0.40 ± 0.08             & \textbf{0.20 ± 0.19}    \\ \hline
\textbf{{AD}}              & PSNR (dB)        & 22.72 ± 3.39        & 23.86 ± 2.16             & 23.16 ± 1.97            & \textbf{30.43 ± 2.15}   \\
                           & SSIM             & 0.85 ± 0.08        & 0.86 ± 0.07             & 0.85 ± 0.08            & \textbf{0.94 ± 0.02}   \\
                           & NRMSE            & 0.41 ± 0.07         & 0.36 ± 0.05              & 0.52 ± 0.07             & \textbf{0.22 ± 0.01}    \\ \hline
\textbf{{RD}}              & PSNR (dB)        & 30.27 ± 2.54        & 30.58 ± 2.57             & 26.22 ± 2.20            & \textbf{30.78 ± 1.65}   \\
                           & SSIM             & 0.94 ± 0.03        & 0.94 ± 0.03             & 0.93 ± 0.03            & \textbf{0.95 ± 0.02}   \\
                           & NRMSE            & 0.18 ± 0.03         & 0.18 ± 0.03              & 0.38 ± 0.09             & \textbf{0.20 ± 0.18}    \\ \hline
                            
\end{tabular}
\end{table*}

\subsection{Comparison with SuperDTI methods}

SuperDTI is a state-of-the-art DTI reconstruction network based on fixed diffusion gradient directions. Although FlexDTI has important advantages in terms of flexibility in the diffusion gradient scheme, it is important to know the performance difference between our method and SuperDTI under the same number of diffusion gradient directions. LLS, BM4D and DIFFnet have also been carried out for comparison.\\
Figure 6 shows the reconstructed results of two estimated tensor-derived variables, i.e., FA and MD from all methods. Compared to SuperDTI, FlexDTI showed a mild performance decline in FA reconstruction (-1.92 dB in PSNR and +0.04 in NRMSE) and was nearly consistent in MD reconstruction. This means that FlexDTI achieves good flexibility without sacrificing too much reconstruction quality.\\
Figure 7 shows the reconstructed results of direction encoded color (DEC) maps from all methods except for DIFFnet, because DIFFnet directly maps diffusion data to FA, MD, AD and RD parameters without quantifying diffusion tensor. Compared to SuperDTI, FlexDTI shows slight differences in the reconstructed DEC maps (-1.75 dB in PSNR and +0.05 in NRMSE), but still maintains good quality.

\begin{figure*}[ht]
\includegraphics[width=30pc,height=22pc]{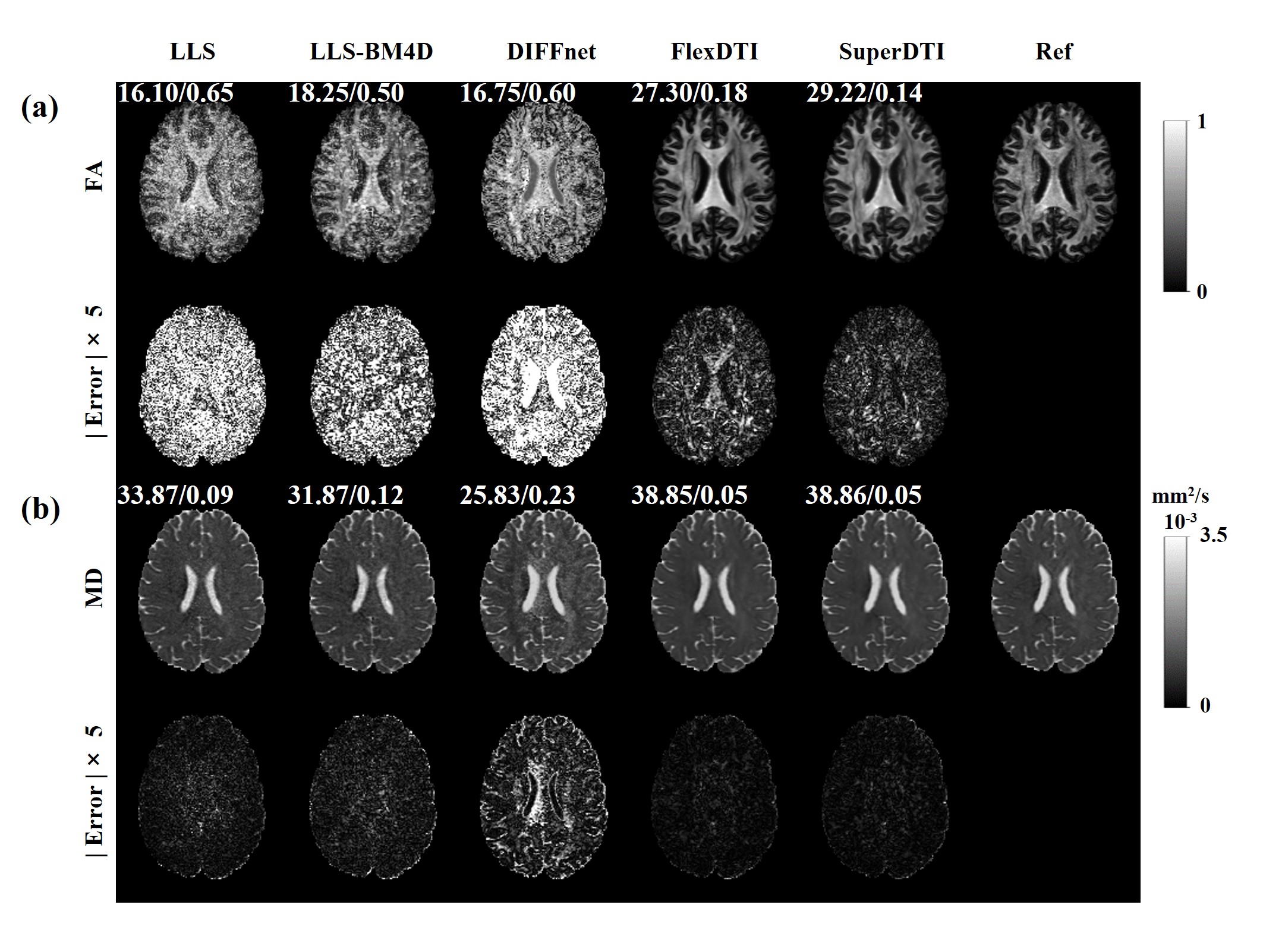}
\centering
\caption{A comparison of FA and MD generated from LLS, LLS-BM4D, DIFFnet, SuperDTI and FlexDTI using six DW images. The diffusion directions were fixed for SuperDTI and flexible for other methods. The references were generated from LLS using 90 DW images. The PSNRs and NRMSEs are given at the upper left corner of each reconstruction image. }\label{Figure 6}
\end{figure*}

\begin{figure*}[ht]
\includegraphics[width=30pc,height=15pc]{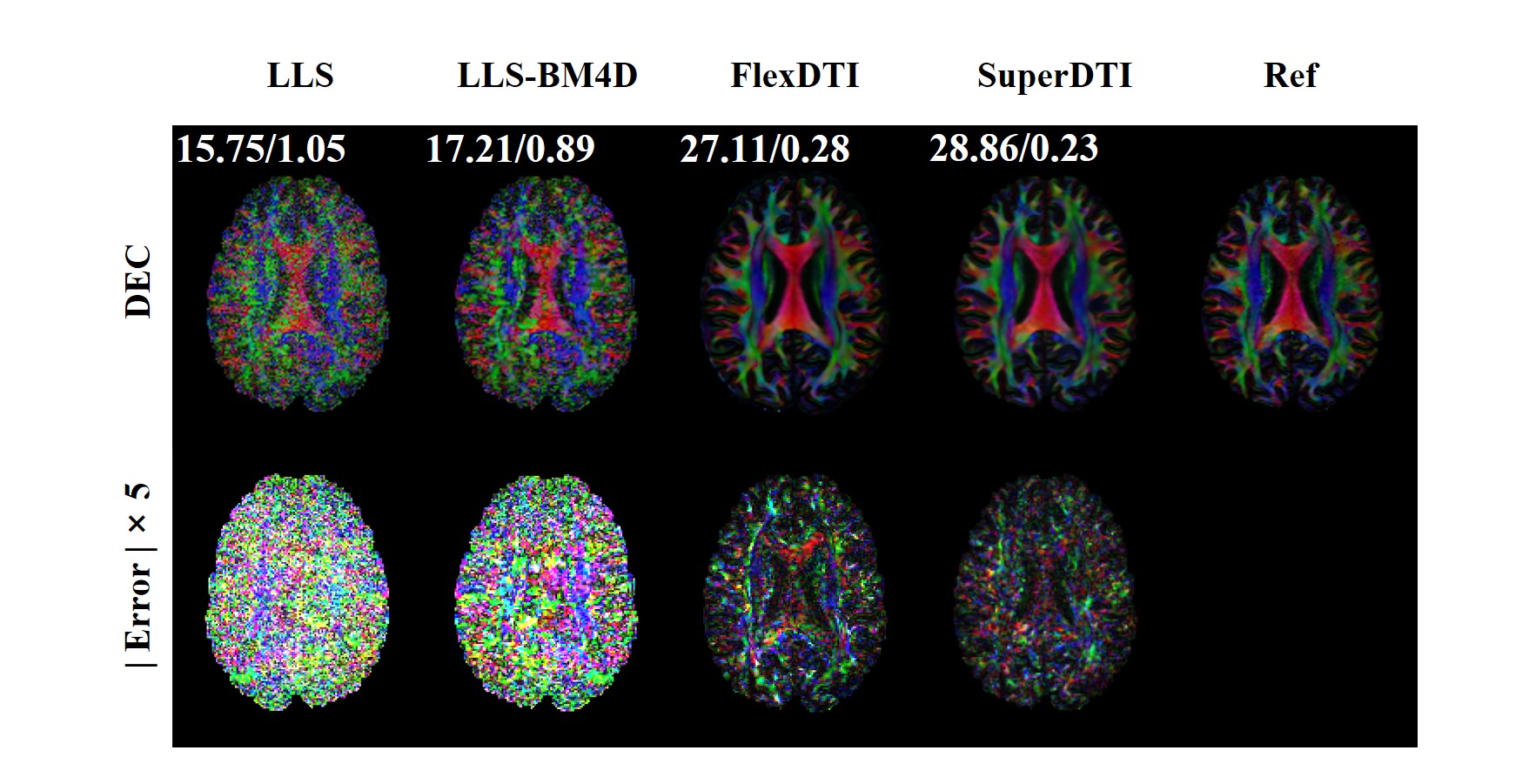}
\centering
\caption{A comparison of DEC maps generated from LLS, LLS-BM4Ds, SuperDTI and FlexDTI using six DW images. The diffusion directions were fixed for SuperDTI and flexible for other methods. The reference was generated from LLS using 90 DW images. The PSNRs and NRMSEs are given at the upper left corner of each reconstructed image. }\label{Figure 7}
\end{figure*}

\section{Discussion}
In this work, FlexDTI was proposed for rapid high-quality DTI reconstruction with flexible diffusion gradient encoding schemes. The input feature maps of the network contain both DW images information and diffusion gradient directions information, allowing the network to learn the mapping relationship between DW images and diffusion tensors for flexible diffusion gradient directions, solving the limitation of previous deep learning methods (Tian \textit{et al} 2020, Li \textit{et al} 2021) that are restricted to a specific diffusion gradient encoding scheme and significantly improving the generalization of the deep learning-based DTI. In addition, the network can reconstruct diffusion tensors from a flexible number of DW images (six or more), thereby enhancing the flexibility. The performance of FlexDTI was systematically evaluated in terms of the quality of tensor-derived parameters, as well as compared to the performance of traditional methods and advanced deep learning methods (Park \textit{et al} 2022) that offer similar levels of flexibility. \\
It is worth pointing out that DIFFnet, another advanced flexible DTI reconstruction method, does not show good enough reconstruction performance. The reason may be twofold. On the one hand, DIFFnet is a pixel-wise reconstruction method, which cannot effectively use the correlation between adjacent pixels. On the other hand, the signals from various diffusion gradient directions are projected onto orthogonal planes to preserve the information of the diffusion gradient directions. However, when the number of diffusion gradient directions decreases, the projection on the plane may result in sparse and scattered signals, making it difficult for the deep network to extract useful features. As a result, the network may require convolution operations with larger receptive fields to capture more global and detailed information.\\
An important design element of the FlexDTI framework is the dynamic convolutional module (Chen \textit{et al} 2020, Zhang \textit{et al} 2021, Huang \textit{et al} 2023), which uses an attention mechanism to construct the convolutional kernel based on the input images. The dynamic convolutional module establishes the correspondence between the position in q-space and the DWI signal, while avoiding the dimensional differences that arise from directly combining the DWI signal with the diffusion gradient direction. In FlexDTI, the dynamic convolution was used to transform diffusion gradient direction information into learnable convolution kernels. This allows for the concurrent encoding of both DWI signal and diffusion gradient direction information into high-dimensional feature maps. This approach enhances the flexibility of network training, moving beyond a straightforward end-to-end mapping strategy. Moreover, in scenarios with limited information, the dynamic convolution module augments the network’s ability to capture crucial information. Compared to other methods, FlexDTI exhibits enhanced extraction of diffusion gradient direction information with a limited number of diffusion gradients, substantiated by the DTI parameters reconstruction results. Dynamic convolution (Jia \textit{et al} 2016) adds only marginal attention computation and kernel construction costs but does not increase the depth or width of the network, rendering the additional computational burden essentially negligible. Moreover, dynamic convolution can be used as a module to replace static convolution in traditional convolutional neural networks, facilitating the implementation of specific network mapping tasks.\\
In the dynamic convolution module, a DW image vector is first concatenated with a specific-length diffusion gradient direction vector, before undergoing parameter generation. If the vector length is too short, a significant loss of information may occur due to global average pooling, resulting in an insufficient contribution of DW image information during parameter generation. Conversely, if the vector length is too long, the information on diffusion gradient directions may become less important since diffusion gradient direction vector is a three-dimensional vector, compromising the network’s encoding ability concerning diffusion gradient directions and ultimately hampering the network’s mapping performance. Thus, vectorizing the DW image to a length of ten strikes the best balance between these competing needs, achieving optimal reconstruction performance.\\
In DTI reconstruction experiments with flexible diffusion gradient directions (Figures 2 and 3), only 6 DW images and 12 DW images were involved respectively. Because clinically, 15-20 DW images (Jones \textit{et al} 2004, Skare \textit{et al} 2000) are typically used for generating DTI parametric maps and using an excessive number of DW images may compromise the purpose of fast DTI.\\ 
The DTI reconstruction of local clinical data (Figure 5) shows that our method is suitable for different scanners, indicating its strong generalization ability. Once the network is trained, it can be used for different diffusion gradient schemes on different types of scanners, which is of great convenience for the clinical application of deep learning-based DTI. It should be pointed out that, for high-resolution DTI, mitigating local volume effects enables finer structural resolution, holding significant implications for enhancing the accuracy of DTI in clinical applications (Wu \textit{et al} 2013, Liu \textit{et al} 2009).  However, due to its prolonged scanning time, current clinical practice often utilizes lower resolution. On a clinical MRI scanner, a more common resolution is a 2×2×2 $mm^3$ acquisition. Figure 8 shows the results of further evaluations of our method on typical clinical brain MRI protocols with three healthy volunteers. Figure 9 shows the performance of FlexDTI in other scan parameters like slice thickness, TE, phase partial Fourier and parallel imaging. According to the quantitative assessment of FA, MD, AD and RD, it can be seen that the proposed method still outperforms other methods for the typical clinical brain MRI protocols and keeps stability under different scan parameters, proving that the proposed model has good generalization performance. 

\begin{figure*}[ht]
\includegraphics[width=30pc,height=48pc]{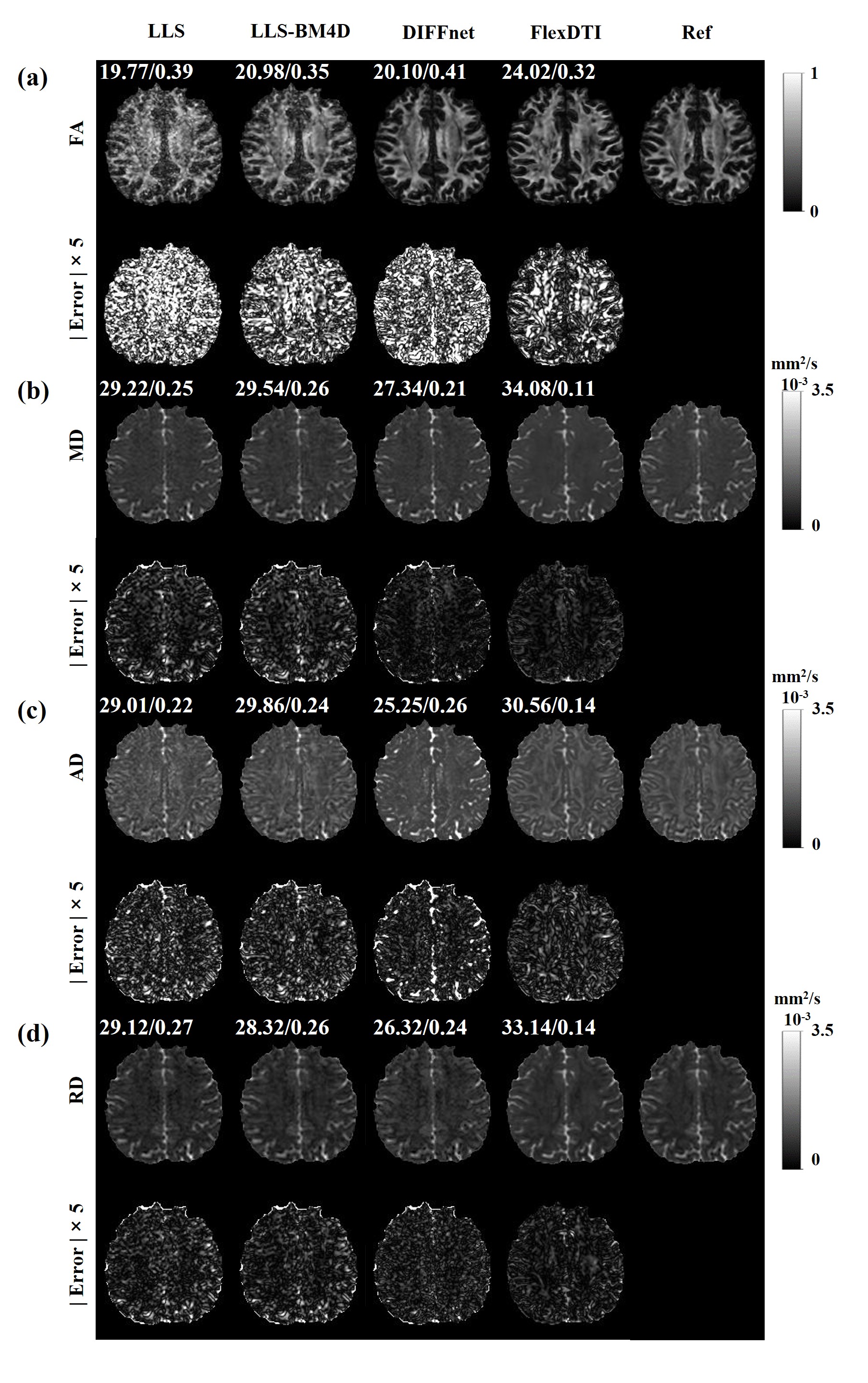}
\centering
\caption{Four parametric maps (FA, MD, AD and RD) reconstructed by LLS, LLS-BM4D, DIFFnet and FlexDTI using 6 DW images with flexible diffusion gradient directions for typical 2×2×2 $mm^3$ acquisition. The references were reconstructed by LLS using 96 DW images. The PSNRs and NRMSEs are given at the upper left corner of each reconstructed image. }\label{Figure 8}
\end{figure*}

\begin{figure*}[ht]
\includegraphics[width=38pc,height=36pc]{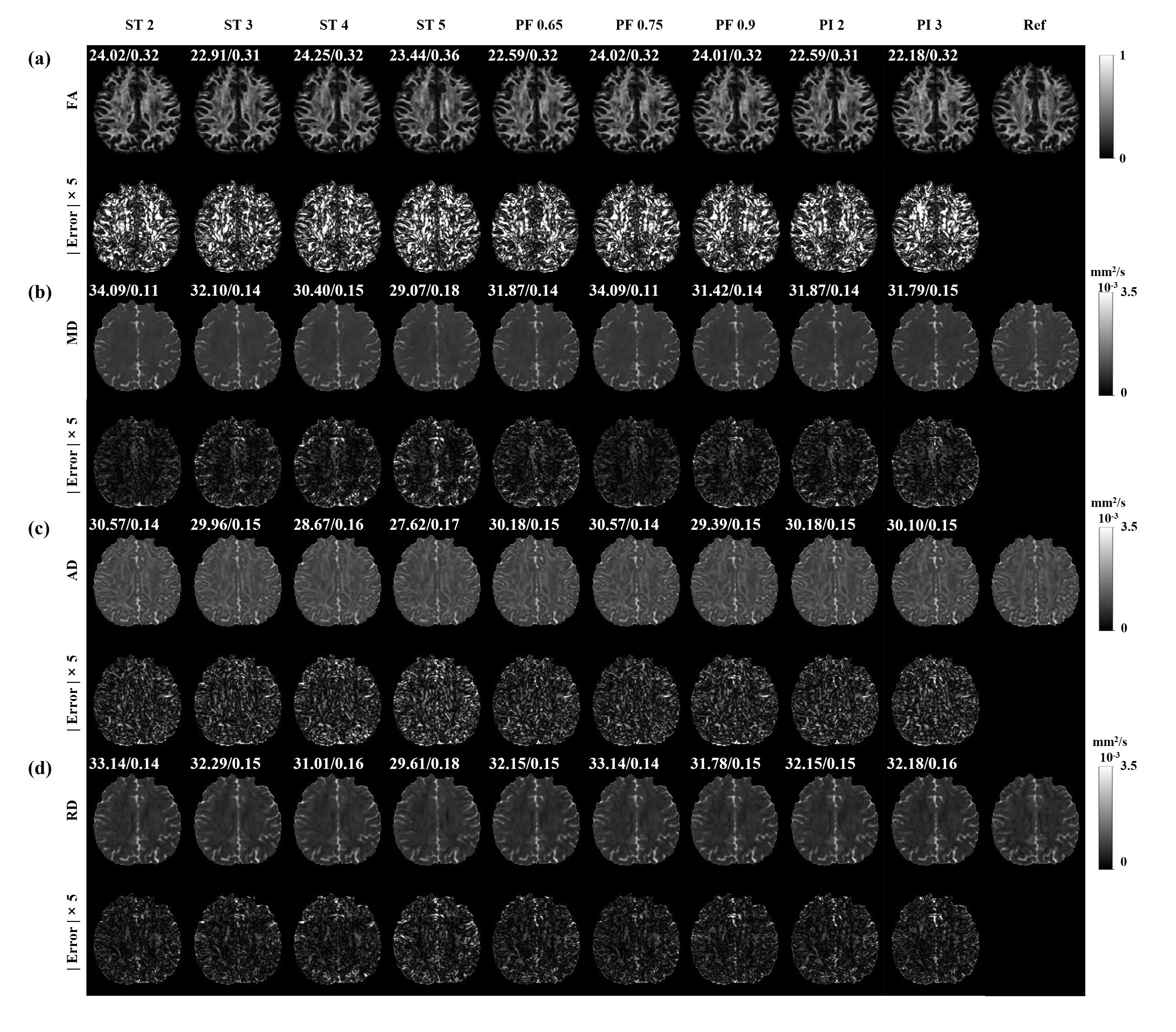}
\centering
\caption{Four parametric maps (FA, MD, AD, and RD) reconstructed by FlexDTI using 6 DW images with flexible diffusion gradient directions under different scan parameters. The references were reconstructed by LLS using 96 DW images. The PSNRs and NRMSEs are given at the upper left corner of each reconstructed image. “ST 2” denotes a slice thickness of 2 mm, “PT 0.65” denotes a phase partial Fourier value of 0.65, “PI 2” denotes an acceleration factor of 2 for parallel imaging and other scan parameters follow a similar convention. }\label{Figure 9}
\end{figure*}

\noindent
In comparison with SuperDTI with fixed diffusion gradient directions (Figure 6), the parameter quantitative result of MD, which is related to the diagonal elements of diffusion tensor, shows that the reconstruction performance of FlexDTI is similar to SuperDTI. The parameter FA, which is related to the off-diagonal elements of diffusion tensor, shows a decrease in PSNR of 1 to 2 dB. Although there were some compromises in the reconstruction quality, the results were acceptable. These observations further demonstrate that FlexDTI has high reconstruction quality while obtaining flexibility. 

\section{Conclusion}
A novel method FlexDTI was proposed to achieve highly efficient diffusion tensor reconstruction with flexible diffusion gradient schemes. The quantitative results show the flexibility and efficiency of this method. By using dynamic convolution to encode the diffusion gradient directions, the network can reconstruct diffusion tensors using DW images with flexible diffusion gradient directions. This solves the problem of traditional deep learning methods that can only reconstruct specific diffusion gradient directions. In addition, we utilized the method of setting the maximum number of input channels of the network to achieve diffusion tensor reconstruction with a flexible number of (six or more) DW images and improved the flexibility of the network. The proposed method meets the requirements of both shortening scan time and increasing flexibility and generalization.

\section{Acknowledgments}
This work was supported in part by the National Key R and D Program of China under Grant 2022YFC2402102, and in part by the National Natural Science Foundation of China under grant numbers 82071913, 12375291 and 22161142024, and the Science and Technology Project of Fujian Province of China under Grant Number 2021Y9154.

\section{Declaration of competing interest}
All authors declare that they have no known competing financial interests or personal relationships that could have appeared to influence the work reported in this paper.

\section{Ethical statement}
The data were obtained from the Human Connectome Project (HCP) WU-Minn-Ox Consortium public database, which was approved by the institutional research ethics committees. For local hospital data, the study protocol was approved by the local institutional research ethics committee, and written informed consents were obtained from the volunteers before the experiments.

\section{Referencing\label{except}}

\smallskip
\begin{harvard}
\item[] Andersson J L R, Skare S and Ashburner J 2003 How to correct susceptibility distortions in spin-echo echo-planar images: Application to diffusion tensor imaging {\it Neuroimage} {\bf 20} 870-88
\end{harvard}
\smallskip

\smallskip
\begin{harvard}
\item[] Auriat A M, Borich M R, Snow N J, Wadden K P and Boyd L A 2015 Comparing a diffusion tensor and non-tensor approach to white matter fiber tractography in chronic stroke {\it NeuroImage Clin.} {\bf 7} 771-81
\end{harvard}
\smallskip

\smallskip
\begin{harvard}
\item[] Cai C B, Wang C, Zeng Y Q, Cai S H, Liang D, Wu Y W, Chen Z, Ding X H and Zhong J H 2018 Single-shot T2 mapping using overlapping-echo detachment planar imaging and a deep convolutional neural network {\it Magn. Reson. Med.} {\bf 80} 2202-14
\end{harvard}
\smallskip

\smallskip
\begin{harvard}
\item[] Caruyer E, Lenglet C, Sapiro G and Deriche R 2013 Design of multishell sampling schemes with uniform coverage in diffusion MRI {\it Magn. Reson. Med.} {\bf 69} 1534-40
\end{harvard}
\smallskip

\smallskip
\begin{harvard}
\item[] Chen Y P, Dai X Y, Liu M C, Chen D D, Yuan L and Liu Z C 2020 Dynamic convolution: Attention over convolution kernels In {\it Proc. IEEE Conf. Comput. Vis. Pattern Recognit. (CVPR)} pp 11027-36
\end{harvard}
\smallskip

\smallskip
\begin{harvard}
\item[] Dabov K, Foi A, Katkovnik V and Egiazarian K 2007 Image denoising by sparse 3-D transform-domain collaborative filtering {\it IEEE Trans. Image Process.} {\bf 16} 2080-95
\end{harvard}
\smallskip

\smallskip
\begin{harvard}
\item[] Edlow B L, Hurwitz S and Edlow J A 2017 Diagnosis of DWI-negative acute ischemic stroke: A meta-analysis {\it Neurology} {\bf 89} 256-62
\end{harvard}
\smallskip

\smallskip
\begin{harvard}
\item[] Elliott R, Zahn R, Deakin J F and Anderson L M 2011 Affective cognition and its disruption in mood disorders {\it Neuropsychopharmacology} {\bf 36} 153-82
\end{harvard}
\smallskip

\smallskip
\begin{harvard}
\item[]Falk T, Mai D, Bensch R, Cicek O, Abdulkadir A, Marrakchi Y, Bohm A, Deubner J, Jackel Z, Seiwald K, Dovzhenko A, Tietz O, Bosco C D, Walsh S, Saltukoglu D, Tay T L, Prinz M, Palme K, Simons M, Diester I, Brox T and Ronneberger O 2019 U-Net: Deep learning for cell counting, detection, and morphometry {\it Nat. Methods} {\bf 16} 67
\end{harvard}
\smallskip

\smallskip
\begin{harvard}
\item[] Glasser M F, Sotiropoulos S N, Wilson J A, Coalson T S, Fischl B, Andersson J L, Xu J, Jbabdi S, Webster M, Polimeni J R, Van Essen D C and Jenkinson M 2013 The minimal preprocessing pipelines for the Human Connectome Project {\it Neuroimage} {\bf 80} 105-24
\end{harvard}
\smallskip

\smallskip
\begin{harvard}
\item[] Golkov V, Dosovitskiy A, Sperl J I, Menzel M I, Czisch M, Samann P, Brox T and Cremers D 2016 Q-space deep learning: Twelve-fold shorter and model-free diffusion MRI Scans {\it IEEE Trans. Med. Imaging} {\bf 35} 1344-51
\end{harvard}
\smallskip

\smallskip
\begin{harvard}
\item[] Huang H T, Yang Q Q, Wang J C, Zhang P J, Cai S H and Cai C B 2023 High-efficient Bloch simulation of magnetic resonance imaging sequences based on deep learning {\it Phys. Med. Biol.} {\bf 68} 085002
\end{harvard}
\smallskip

\smallskip
\begin{harvard}
\item[] Jia X, De Brabandere B, Tuytelaars T and Gool L V 2016 Dynamic filter networks {\it Advances in Neural Information Processing Systems 29 (NIPS 2016)} vol 29, pp 667–75
\end{harvard}
\smallskip

\smallskip
\begin{harvard}
\item[] Jones D K 2004 The effect of gradient sampling schemes on measures derived from diffusion tensor MRI: A Monte Carlo study {\it Magn. Reson. Med.} {\bf 51} 807-15
\end{harvard}
\smallskip

\smallskip
\begin{harvard}
\item[] Jovicich J, Czanner S, Greve D, Haley E, Van Der Kouwe A, Gollub R, Kennedy D, Schmitt F, Brown G, MacFall J, Fischl B and Dale A 2006 Reliability in multi-site structural MRI studies: Effects of gradient non-linearity correction on phantom and human data  {\it Neuroimage} {\bf 30} 436-43
\end{harvard}
\smallskip

\smallskip
\begin{harvard}
\item[] Kono K, Inoue Y, Nakayama K, Shakudo M, Morino M, Ohata K, Wakasa K and Yamada R 2001 The role of diffusion-weighted imaging in patients with brain tumors {\it Am. J. Neuroradiol.} {\bf 22} 1081-88
\end{harvard}
\smallskip

\smallskip
\begin{harvard}
\item[] Li H Y, Liang Z F, Zhang C Y, Liu R Y, Li J, Zhang W H, Liang D, Shen B W, Zhang X L, Ge Y L, Zhang J Y and Ying L 2021 SuperDTI: Ultrafast DTI and fiber tractography with deep learning {\it Magn. Reson. Med.} {\bf 86} 3334-47
\end{harvard}
\smallskip

\smallskip
\begin{harvard}
\item[] Li Z W, Gong T, Lin Z C, He H J, Tong Q Q, Li C, Sun Y, Yu F and Zhong J H 2019 Fast and robust diffusion kurtosis parametric mapping using a three-dimensional convolutional neural network  {\it IEEE Access} {\bf 7} 71398-411
\end{harvard}
\smallskip

\smallskip
\begin{harvard}
\item[] Liu S N, Li H X, Liu Y Y, Cheng G X, Yang G, Wang H F, Zheng H R, Liang D and Zhu Y J 2022 Highly accelerated MR parametric mapping by undersampling the k-space and reducing the contrast number simultaneously with deep learning {\it Phys. Med. Biol.} {\bf 67} 185004
\end{harvard}
\smallskip

\smallskip
\begin{harvard}
\item[] Liu X X, Zhu T, Gu T L and Zhong J H 2009 A practical approach to in vivo high-resolution diffusion tensor imaging of rhesus monkeys on a 3-T human scanner {\it Magn. Reson. Imaging} {\bf 27} 335-46
\end{harvard}
\smallskip

\smallskip
\begin{harvard}
\item[] Moseley M E, Cohen Y, Mintorovitch J, Chileuitt L, Shimizu H, Kucharczyk J, Wendland M F and Weinstein P R 1990 Early detection of regional cerebral ischemia in cats: Comparison of diffusion- and T2-weighted MRI and spectroscopy {\it Magn. Reson. Med.} {\bf 14} 330-46
\end{harvard}
\smallskip

\smallskip
\begin{harvard}
\item[] Muller H P, Turner M R, Grosskreutz J, Abrahams S, Bede P, Govind V, Prudlo J, Ludolph A C, Filippi M and Kassubek J 2016 A large-scale multicentre cerebral diffusion tensor imaging study in amyotrophic lateral sclerosis {\it J. Neurol. Neurosurg. Psychiatry} {\bf 87} 570-79
\end{harvard}
\smallskip

\smallskip
\begin{harvard}
\item[] Narayana P A, Yu X T, Hasan K M, Wilde E A, Levin H S, Hunter J V, Miller E R, Patel V K S, Robertson C S and Mccarthy J J 2015 Multi-modal MRI of mild traumatic brain injury {\it NeuroImage Clin.} {\bf 7} 87-97
\end{harvard}
\smallskip

\smallskip
\begin{harvard}
\item[] Park J, Jung W, Choi E J, Oh S H, Jang J, Shin D, An H and Lee J 2022 DIFFnet: Diffusion parameter mapping network generalized for input diffusion gradient schemes and b-value {\it IEEE Trans. Med. Imaging} {\bf 41} 491-99
\end{harvard}
\smallskip

\smallskip
\begin{harvard}
\item[] Pasternak O, Kelly S, Sydnor V J and Shenton M E 2018 Advances in microstructural diffusion neuroimaging for psychiatric disorders {\it Neuroimage} {\bf 182} 259-82
\end{harvard}
\smallskip

\smallskip
\begin{harvard}
\item[] Rive M M, Van Rooijen G, Veltman D J, Phillips M L, Schene A H and Ruhe H G 2013 Neural correlates of dysfunctional emotion regulation in major depressive disorder. A systematic review of neuroimaging studies {\it Neurosci. Biobehav. Rev.} {\bf 37} 2529-53
\end{harvard}
\smallskip

\smallskip
\begin{harvard}
\item[] Rovaris M and Filippi M 2007 Diffusion tensor MRI in multiple sclerosis {\it J. Neuroimaging} {\bf 17} 27S-30S
\end{harvard}
\smallskip

\smallskip
\begin{harvard}
\item[] Skare S, Hedehus M, Moseley M E and Li T Q 2000 Condition number as a measure of noise performance of diffusion tensor data acquisition schemes with MRI {\it J. Magn. Reson.} {\bf 147} 340-52
\end{harvard}
\smallskip

\smallskip
\begin{harvard}
\item[] Song S K, Sun S W, Ramsbottom M J, Chang C, Russell J and Cross A H 2002 Dysmyelination revealed through MRI as increased radial (but unchanged axial) diffusion of water {\it Neuroimage} {\bf 17} 1429-36
\end{harvard}
\smallskip

\smallskip
\begin{harvard}
\item[] Sotiropoulos S N, Jbabdi S, Xu J, Andersson J L, Moeller S, Auerbach E J, Glasser M F, Hernandez M, Sapiro G, Jenkinson M, Feinberg D A, Yacoub E, Lenglet C, Ven Essen D C, Ugurbil K and Behrens T E J 2013 Advances in diffusion MRI acquisition and processing in the Human Connectome Project {\it Neuroimage} {\bf 80} 125-43
\end{harvard}
\smallskip

\smallskip
\begin{harvard}
\item[] Stadnik T W, Chaskis C, Michotte A, Shabana W M, Van Rompaey K, Luypaert R, Budinsky L, Jellus V and Osteaux M 2001 Diffusion-weighted MR imaging of intracerebral masses: Comparison with conventional MR imaging and histologic findings {\it Am. J. Neuroradiol.} {\bf 22} 969-76
\end{harvard}
\smallskip

\smallskip
\begin{harvard}
\item[] Stejskal E O and Tanner J E 1965 Spin diffusion measurements: Spin echoes in the presence of a time dependent field gradient {\it J. Chem. Phys.} {\bf 42} 288-92
\end{harvard}
\smallskip

\smallskip
\begin{harvard}
\item[] Szczepankiewicz F, Lasic S, Van Westen D, Sundgren P C, Englund E, Westin C F, Stahlberg F, Latt J, Topgaard D and Nilsson Markus 2015 Quantification of microscopic diffusion anisotropy disentangles effects of orientation dispersion from microstructure: Applications in healthy volunteers and in brain tumors {\it Neuroimage} {\bf 104} 241-52
\end{harvard}
\smallskip

\smallskip
\begin{harvard}
\item[] Tae W S, Ham B J, Pyun S B, Kang S H and Kim B J 2018 Current clinical applications of diffusion-tensor imaging in neurological disorders {\it J. Clin. Neurol.} {\bf 14} 129-40
\end{harvard}
\smallskip

\smallskip
\begin{harvard}
\item[] Tetreault P, Harkins K D, Baron C A, Stobbe R, Does M D and Beaulieu C 2020 Diffusion time dependency along the human corpus callosum and exploration of age and sex differences as assessed by oscillating gradient spin-echo diffusion tensor imaging {\it Neuroimage} {\bf 210} 116533
\end{harvard}
\smallskip

\smallskip
\begin{harvard}
\item[] Tian Q, Bilgic B, Fan Q, Liao C, Ngamsombat C, Hu Y, Witzel T, Setsompop K, Polimeni J R and Huang S Y 2020 DeepDTI: High-fidelity six-direction diffusion tensor imaging using deep learning {\it Neuroimage } {\bf 219} 117017
\end{harvard}
\smallskip

\smallskip
\begin{harvard}
\item[] Van der Walt A, Kolbe S C, Wang Y E, Klistorner A, Shuey N, Ahmadi G, Paine M, Marriott M, Mitchell P, Egan G F, Butzkueven H and Kilpatrick T J 2013 Optic nerve diffusion tensor imaging after acute optic neuritis predicts axonal and visual outcomes {\it PLoS One} {\bf 8} e83825
\end{harvard}
\smallskip

\smallskip
\begin{harvard}
\item[] Van Essen D C, Ugurbil K, Auerbach E, Sotiropoulos S N, Jbabdi S, Xu J, Andersson J L, Moeller S, Auerbach E J, Glasser M F, Hernandez M, Sapiro G, Jenkinson M, Feinberg D A, Yacoub E, Lenglet C, Van Essen D C, Ugurbil K, Behrens T E and Consortium W U-M H 2013 The Human Connectome Project: A data acquisition perspective {\it Neuroimage } {\bf 62} 2222-31
\end{harvard}
\smallskip

\smallskip
\begin{harvard}
\item[] Wang J C, Geng W H, Wu J, Kang T S, Wu Z G, Lin J Z, Yang Y, Cai C B and Cai S H 2023 Intravoxel incoherent motion magnetic resonance imaging reconstruction from highly under-sampled diffusion-weighted PROPELLER acquisition data via physics-informed residual feedback unrolled network {\it Phys. Med. Biol.} {\bf 68} 175022
\end{harvard}
\smallskip

\smallskip
\begin{harvard}
\item[] Wu D, Xu J, McMahon M T, Van Zijl P C M, Mori S, Northington F J and Zhang J 2013 In vivo high-resolution diffusion tensor imaging of the mouse brain {\it Neuroimage } {\bf 83} 18-26
\end{harvard}
\smallskip

\smallskip
\begin{harvard}
\item[] Yang Q Q, Lin Y H, Wang J C, Bao J F, Wang X Y, Ma L C, Zhou Z H, Yang Q Z, Cai S H, He H J, Cai C B, Dong J Y, Cheng J L, Chen Z, Zhong J H 2022 MOdel-based synthetic data-driven learning (MOST-DL): Application in single-shot T2 mapping with severe head motion using overlapping-echo acquisition {\it IEEE Trans. Med. Imaging} {\bf 41} 3167-81
\end{harvard}
\smallskip

\smallskip
\begin{harvard}
\item[] Zhang H, Schneider T, Wheeler Kingshott C A and Alexander DC 2012 NODDI: Practical in vivo neurite orientation dispersion and density imaging of the human brain {\it Neuroimage } {\bf 61} 1000-16
\end{harvard}
\smallskip

\smallskip
\begin{harvard}
\item[] Zhang J, Wu J, Chen S J, Zhang Z Y, Cai S H, Cai C B and Chen Z 2019 Robust single-shot T2 mapping via multiple overlapping-echo acquisition and deep neural network {\it IEEE Trans. Med. Imaging} {\bf 38} 1801-11
\end{harvard}
\smallskip

\smallskip
\begin{harvard}
\item[] Zhang J P, Xie Y T, Xia Y and Shen C H 2021 DoDNet: Learning to segment multi-organ and tumors from multiple partially labeled datasets In {\it Proc. IEEE Conf. Comput. Vis. Pattern Recognit. (CVPR)} pp 1195-204
\end{harvard}
\smallskip

\smallskip
\begin{harvard}
\item[] Zhang R Z, Zhao L, Lou W T, Abrigo J M, Mok V C T, Chu W C W, Wang D F and Shi L 2018 Automatic segmentation of acute ischemic stroke from DWI using 3-D fully convolutional DenseNets {\it IEEE Trans. Med. Imaging} {\bf 37} 2149-60
\end{harvard}
\smallskip

\smallskip
\begin{harvard}
\item[] Zhang X L, Guo D, Huang Y M, Chen Y, Wang L S, Huang F, Xu Q and Qu X B 2020 Image reconstruction with low-rankness and self-consistency of k-space data in parallel MRI {\it Med. Image Anal.} {\bf 63} 101687
\end{harvard}
\smallskip

\end{document}